\documentclass[twoside,a4paper,11pt]{sea10}
\usepackage{graphicx}
\begin{document}
\pagenumbering{arabic}
\pagestyle{myheadings}
\thispagestyle{empty}
{\flushleft\includegraphics[width=\textwidth,bb=58 650 590 680]{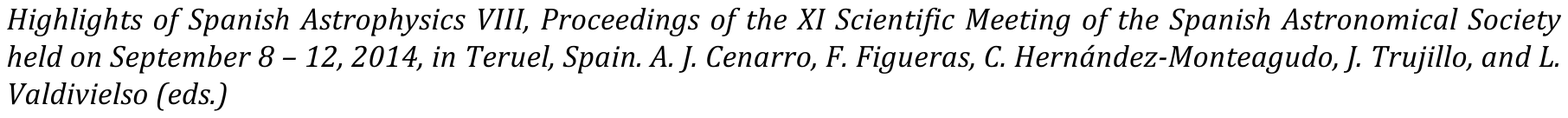}}
\vspace*{0.2cm}
\begin{flushleft}
{\bf {\LARGE
%
Selection of AGN candidates in the GOODS-South Field through SPITZER/MIPS 24 microns variability 
%
}\\
\vspace*{1cm}
%
J. Garc\'ia-Gonz\'alez$^{1}$,
A. Alonso-Herrero$^{1,2}$, 
P. G.~P\'erez-Gonz\'alez$^3$,
A. Hern\'an-Caballero$^1$,
V.~L.~Sarajedini$^4$,
and 
V.~Villar$^3$
%
}\\
\vspace*{0.5cm}
%

$^1$ Instituto de F\'isica de Cantabria, CSIC-UC, 39005 Santander, Spain\\
$^2$ Augusto Gonz\'alez Linares Senior Research Fellow\\
$^3$ Departamento de Astrof\'isica, UCM, 28040 Madrid, Spain\\
$^4$ Departament of Astronomy, University of Florida, Gainesville, FL 32611, USA

%
\end{flushleft}
%
\markboth{
AGN candidates selected by MIPS 24~$\mu$m variability
}{ 
%
Garc\'ia-Gonz\'alez et al.
%
}
\thispagestyle{empty}
\vspace*{0.4cm}
\begin{minipage}[l]{0.09\textwidth}
\ 
\end{minipage}
\begin{minipage}[r]{0.9\textwidth}
\vspace{1cm}
\section*{Abstract}{\small
%
 We present a study of galaxies showing mid-infrared variability in the deepest 
 {\em Spitzer/MIPS} 24~$\mu$m surveys in the GOODS-South field. We divide the dataset 
 in epochs and subepochs to study the long-term (months-years) and the short-term (days) 
 variability. We use a $\chi^2$-statistics method to select AGN candidates with a 
 probability $\leq$ 1\% that the observed variability is due to statistical errors alone.
 We find 39 (1.7\% of the parent sample) sources that show long-term variability and
 55 (2.2\% of the parent sample) showing short-term variability. We compare our candidates 
 with AGN selected in the X-ray and radio  bands, and AGN candidates selected by their IR
 emission. Approximately, 50\% of the MIPS 24 $\mu$m variable sources would be identified
 as AGN with these other methods. Therefore, MIPS 24 $\mu$m variability is a new method 
 to identify AGN candidates, possibly dust obscured and low luminosity AGN that might be
 missed by other methods. However, the contribution of the MIPS 24 $\mu$m 
variable identified AGN to the general AGN population is small ($\leq$ 13\%) in GOODS-South. 
%
\normalsize}
\end{minipage}
%
%
%
\section{Introduction \label{intro}}

Variability can be used to select active galactic nucleus (AGN). Practically all AGN vary
on time-scales from hours to millions of years (\cite{UMU}; \cite{Hickox2014}). Any variability
detected in galaxies on human time-scales must originate in the
nuclear region, because the typical timescale for star formation variability is $\geq$ 100 Myr 
(\cite{Hickox2014}). In particular low-luminosity AGN are expected to
  show stronger 
variability than the luminous ones (\cite{Tre}). Therefore, 
  variability is likely to be an effective method to select
low-luminosity  AGN. Although the 
mechanisms that produce variability are not well understood, the main
explanations involve disk instabilities (\cite{Pe}) or changes in
the amount of accreting material (\cite{HB}).  

The aim of this work is to identify AGN through mid-IR
variability in the GOODS-South field  using 24~$\mu$m 
observations taken with  the Multiband Imaging Photometer for {\it
  Spitzer} (MIPS, \cite{Rie}) 
on board the {\em Spitzer} Space Telescope (\cite{We}).   
The near and mid-IR nuclear emission of AGN,  once the
  stellar component is subtracted, is believed to be due to 
hot and warm dust ($200-2000\,$K) in the dusty torus of the AGN, according to the 
  Unified Model \cite{Antonucci1993}. In this context, variability in the accretion
disk emission would cause delayed variability in the near and mid-IR as the hot and
warm dust, respectively, in the torus react to this change (see \cite{HK} and references therein).
Our choice of using mid-IR variability allows a novel way to select low luminosity and possibly
  obscured AGN that might be otherwise missed by other techniques. Apart
  from this work, there is only other IR variability study in the
  Bo\"otes cosmological field using IRAC data (\cite{Ko}). They used the most sensitive IRAC bands at 3.6 and 4.5 
$\mu$m and found that 1.1\% of the sources
satisfied their variability criteria.

\section{The data and photometry}

We compiled all the data taken around the GOODS-South field with the
MIPS instrument at 24~$\mu$m by querying the {\em Spitzer} Heritage
Archive. This field was
observed by {\em Spitzer} during several campaigns from January 2004
to March 2007. We focused our study on a region around
RA=$3^h32^m36^s$ (J2000) and DEC=$-27^o 48'39''$
(J2000). We divided these data sets into 7 different epochs in order to detect
variable sources. For this study we decided to exclude Epochs 2, 4, and 5 because their
  FoV is small when compared to the other epochs. The common area for the epochs 1, 3, 6, and 7 is
$\sim$1360~arcmin$^2$. They probe time scales of months up to
  three years,
  and henceforth are used to study the long-term variability covering
  a period of over three years. We also
subdivided Epoch 7 in three epochs, namely Epochs 7a, 
7b, and 7c to study the short-term 
variability. The short-term variability epochs have a common area of
$\sim$1960~arcmin$^2$ and probe time scales of days, covering a
  period of 7 days.  

To study the temporal variability of MIPS $24\,\mu$m sources detected
in the common regions we built a source catalog for each epoch
  and subepoch. We used SExtractor 
(Source-Extractor, \cite{BA}) to detect sources and the Image
Reduction and Analysis Facility (IRAF) to perform the photometry following
the procedure explained in \cite{Perez-Gonzalez2005} and \cite{Perez-Gonzalez2008}.
 We obtained a 24~$\mu$m
source catalog for each epoch. In this work we restrict the
analysis to sources above the 5$\sigma$ detection limit in the shallowest data in the mosaics. This corresponds to MIPS
  $24\,\mu$m fluxes of $80\,\mu$Jy and $100\,\mu$Jy for the long-term
  and the short-term epochs, respectively.  We also discarded sources with neighbours at distances of less than 10'' 
to minimize crowding effects in the photometry that could affect the flux measurements and produce false variability positives. 
To identify the common sources in all the epochs we cross-matched the
catalogs using a 2'' radius, imposing additionally that
  the 2'' criterion was fulfilled in each pair of epochs.  Our final catalogs contain 2277 
MIPS 24 $\mu$m sources detected in  Epochs 1, 3, 6, and 7 and 2452 MIPS 24 $\mu$m sources in Epochs 7a, 7b, and 7c,
covering an area of 1360 and 1960 arcmin$^2$, respectively. 

\section{Selection of MIPS $24\,\mu$m variable sources}

To select the 24 $\mu$m variable sources we used a $\chi^2$-statistics method to account for
 the variations of intrinsic flux uncertainties of each epoch (related to differences in depth). 
This is the case for our study as different epochs have different depths and within a given mosaic 
there are some variations in depth. The latter effect is most prominent in epoch 7, which is deeper in the center.
This method associates each flux with its error. The $\chi^2$-statistics is defined as 
follows: $\chi^2=\sum_{i=1}^n{(F_i-\bar{F})^2\over{\sigma_i^2}}$, where n is the number of 
epochs, $F_i$ is the flux in a given epoch, $\sigma_i$ is the associated error in the $i^{th}$ epoch, and $\bar{F}$ is the mean flux.

We calculated the $\chi^2$ value for each source without neighbours.  We selected as variable candidates
 those sources above the 99$^{th}$ percentile of the $\chi^2$ distribution expected from photometric errors
 alone. That is, only 1\% of non-variable sources satisfy the selection criteria. This
value corresponds to $\chi^2\geq11.34$ for the 4 epochs sample (3
degrees of freedom) and $\chi^2\geq9.21$ for the 3 epochs one (2
degrees of freedom).

Every object with a $\chi^2$ value higher than the threshold was
visually inspected to remove artefacts. We also discarded objects that
fell close to the edge of the mosaic. After discarding problematic objects,  our final
  sample contains 39
MIPS $24\,\mu$m long-term variable sources (0.03 sources $\times$ arcmin$^{-2}$) and 55 MIPS
$24\,\mu$m short-term 
variable sources (0.03 sources $\times$ arcmin$^{-2}$). The selected MIPS $24\,\mu$m long-term and short-term variable
  sources represent 1.7 and 2.2\% of the original parent samples, 
respectively. Only two sources are identified as having both, long and short-term 
variability. 28 MIPS $24\,\mu$m long-term and 33 MIPS $24\,\mu$m short-term are located in the Extended Chandra Deep Field South (E-CDFS).

\begin{figure}
\center
\includegraphics[width=60mm]{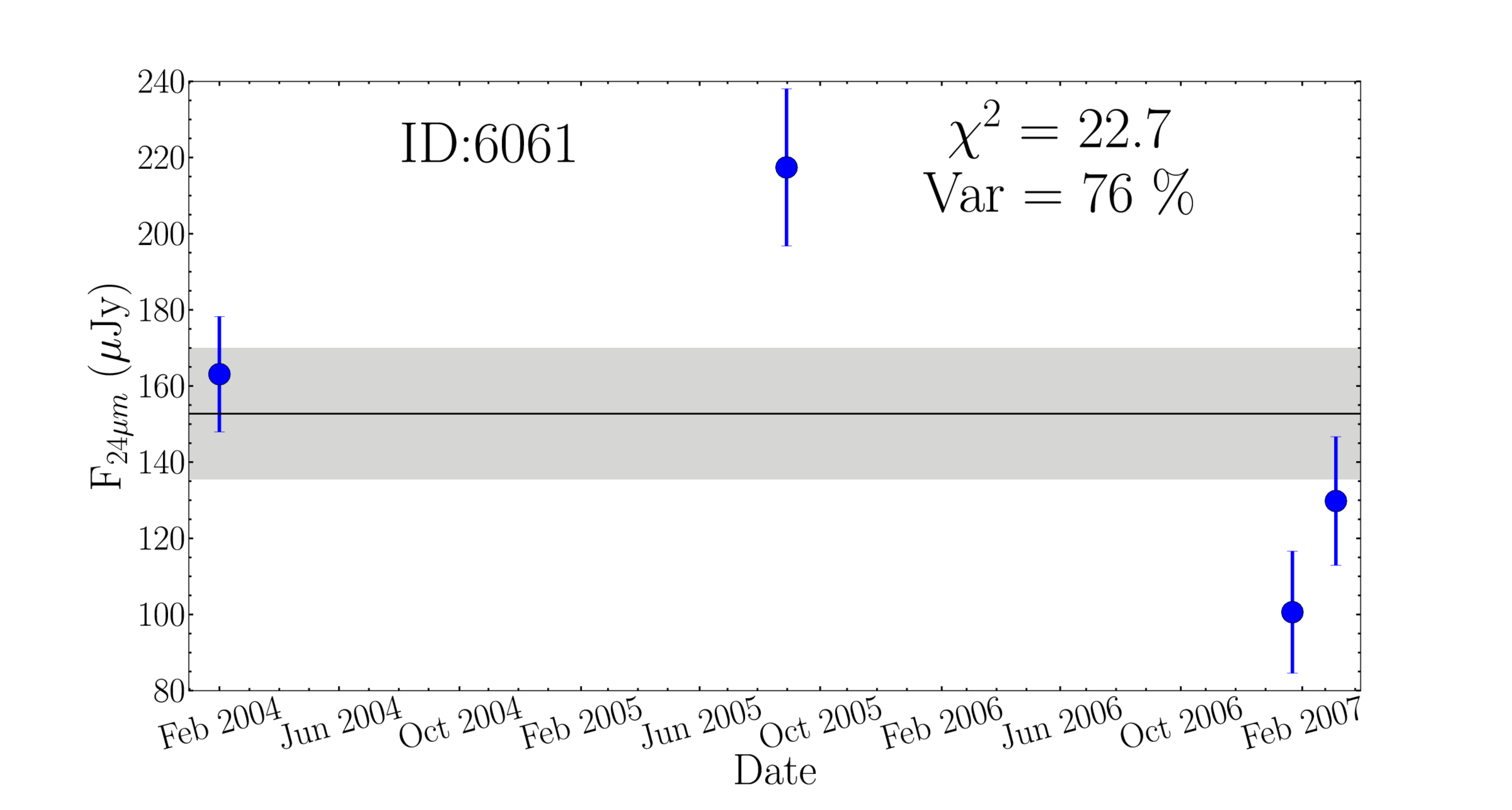} 
\includegraphics[width=60mm]{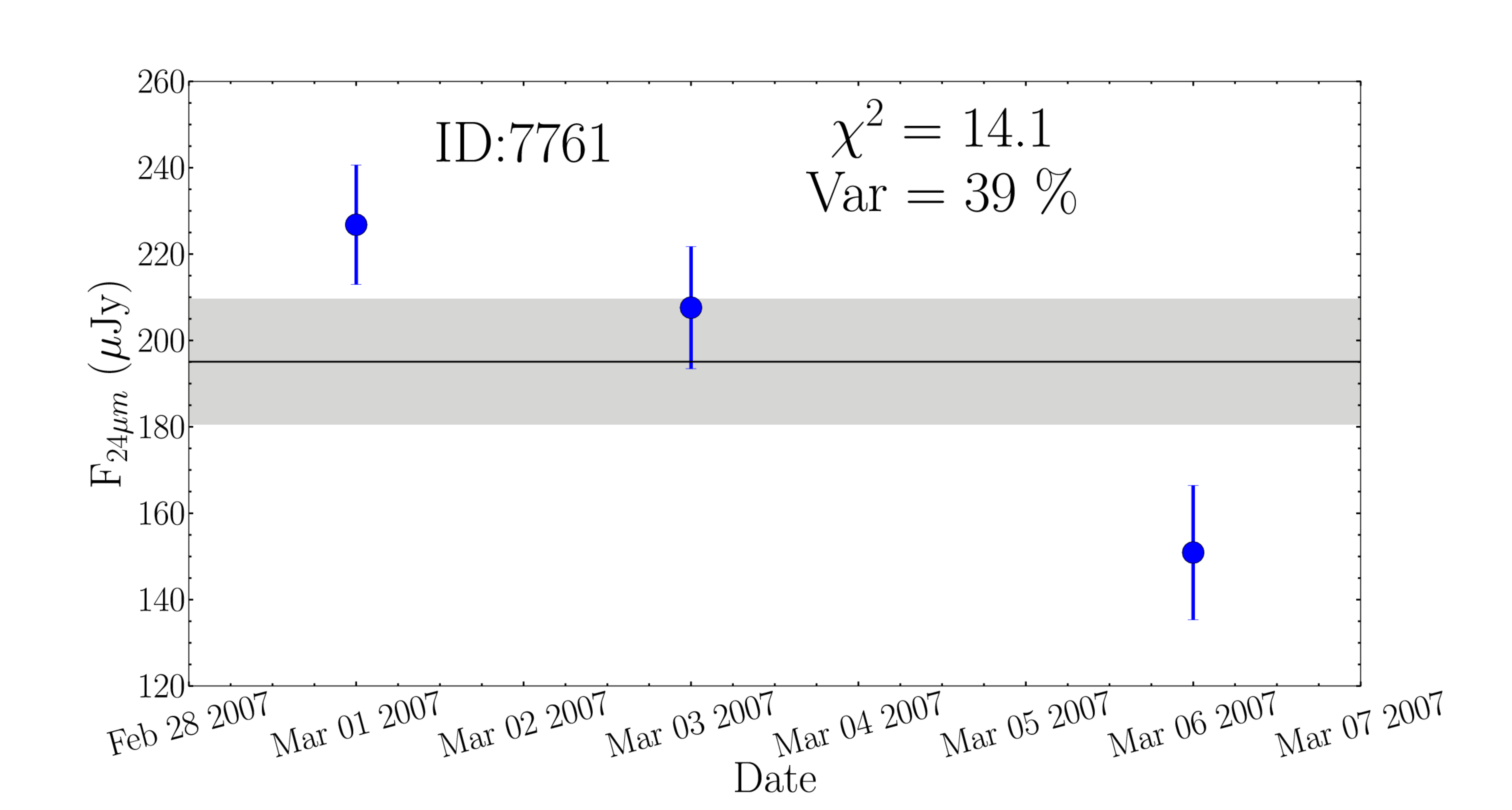} 

\caption{\label{light-curves} Examples of light curves of two MIPS 24~$\mu$m variable
      sources in GOODS-South. The left panel corresponds to a
      long-term variable source (four epochs) and the right
      panel to a short-term variable candidate (three epochs). The
      flux for each epoch is plotted with its corresponding
      photometric error. The solid line is the 24~$\mu$m mean flux of
      the source and the gray shaded area is the average of the errors
      of the source. Each plot lists the name of the source, the
      $\chi^2$ value, and Var.
}
\end{figure}

\section{Properties of the MIPS 24 $\mu$m variable sources}

 The $24\,\mu$m fluxes of the variable sources are
  dominated by sources with mean fluxes below 300~$\mu$Jy. The
median 24 $\mu$m flux is 168~$\mu$Jy for the long-term variable
sources and 209~$\mu$Jy 
for the short-term variable sources. This slight difference in the
median values of the 24~$\mu$m fluxes for long and short-term
variability is likely reflecting the different depths (i.e., 5$\sigma$
  detection limits) of the epochs
rather than different intrinsic properties of the sources).

An estimate of the variability is the ratio between the maximum
and minimum values and the mean flux $\bar f$ measured as a \%: $Var={{f_{\rm max}-f_{\rm min}}\over{\bar f}} \times 100$. The 
typical 24~$\mu$m $Var$ values of the long-term and short-term variable sources are
37-43\%, with typical errors of 12-13\%. In Figure \ref{light-curves} we show two example light curves, one of
long-term and the other of short-term variable sources. Each plot
shows the name of the source, the $\chi^2$ value and the  measure
  of the variability $Var$).

\begin{figure}
\center
  \includegraphics[height=62mm]{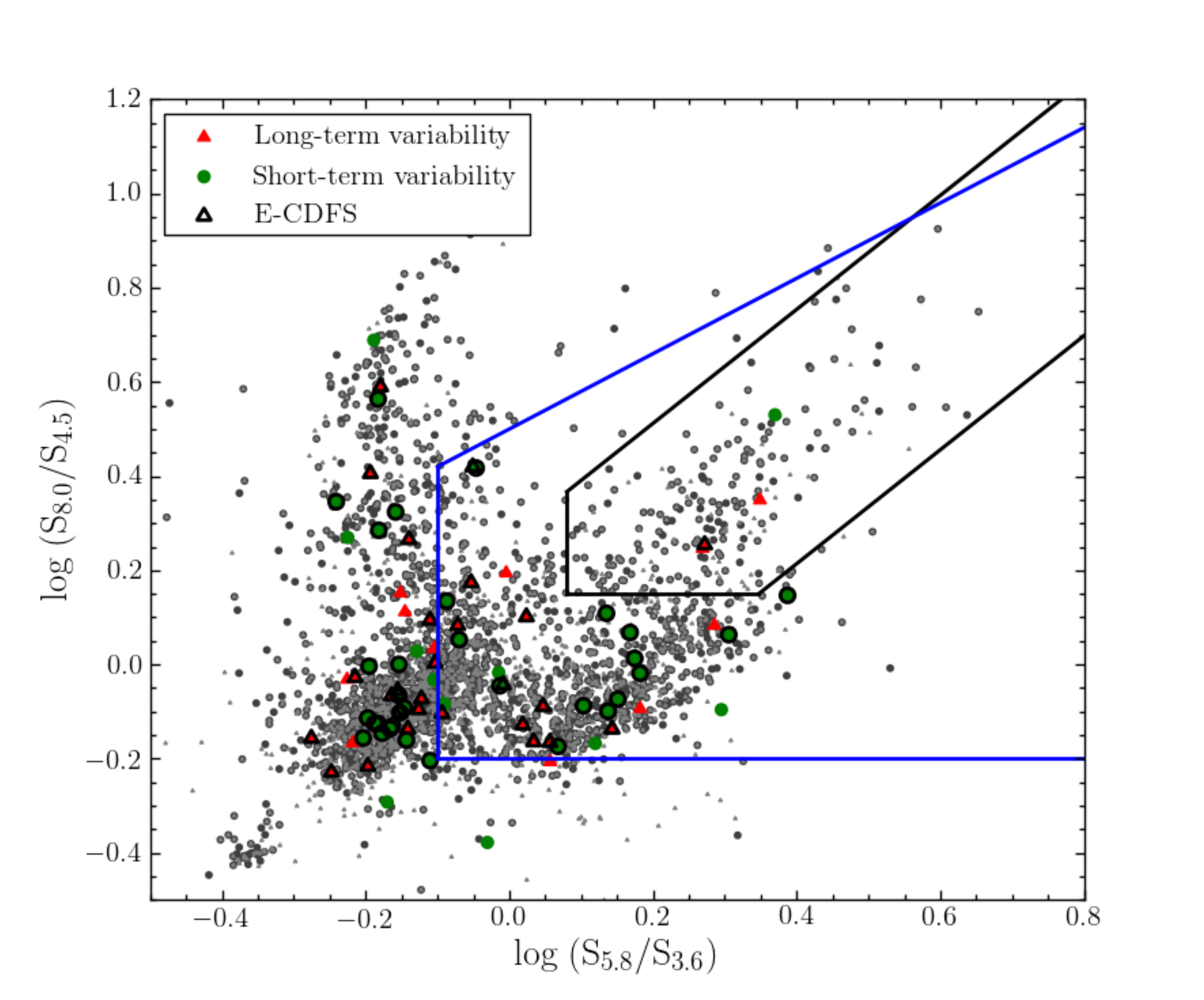}
  \includegraphics[height=62mm]{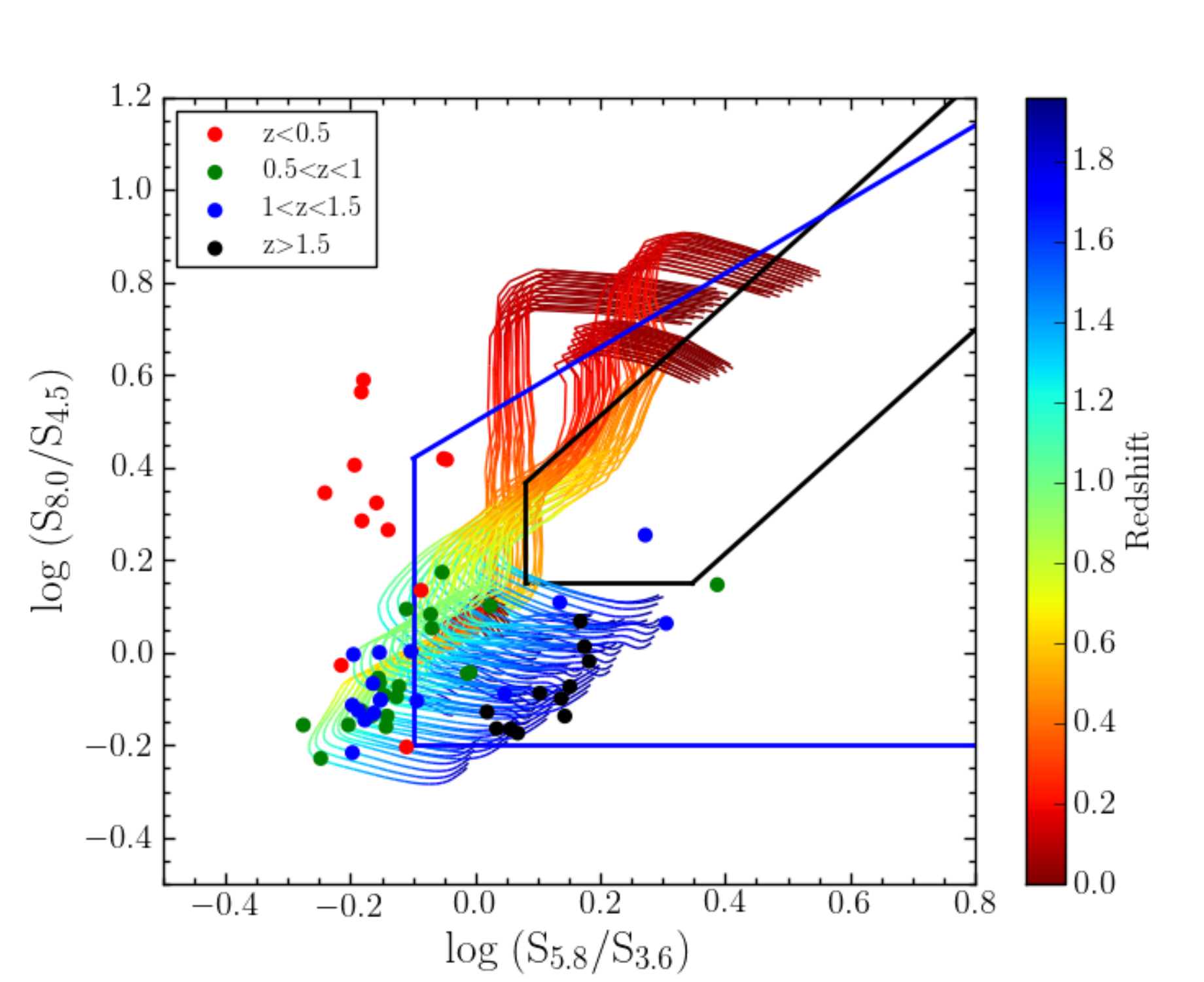}

  \caption{\label{diagrama-color-color}IRAC colour-colour plot of MIPS $24\,\mu$m variable sources in GOODS-South 
from the {\it Rainbow} database (left panel) and plotted according to their redshift (right panel). The different 
AGN wedges are shown as blue solid line for
\cite{La} and black solid line for \cite{Do}. The multicoloured lines (right panel) are the predicted IRAC colours
of the star-forming templates for four different templates from \cite{Do} with a 20\% AGN contribution.}  
\end{figure}

We also studied the Spitzer-IRAC mid-IR (3.6, 4.5, 5.8, and 8.0 $\mu$m) properties of 
the MIPS $24\,\mu$m variable sources as the IRAC emission has also been
used to select AGN candidates (e.g., \cite{La}; \cite{Ste};
\cite{Almu}; \cite{Do}; \cite{La13}). To obtain the IRAC data for our sources, we used
the {\it Rainbow} Cosmological Surveys Database, which contains
multi-wavelength  
photometric data as well as spectroscopic information for sources in different cosmological fields, including 
GOODS-South \cite{Perez-Gonzalez2005,Perez-Gonzalez2008}. In Figure \ref{diagrama-color-color} we show the IRAC
colour-colour plot. 44\% of the variable sources fall in the \cite{La} AGN wedge. The majority of the variable
sources are compatible with a low AGN contribution in the IR.

From the {\it Rainbow} database we also obtained the photometric redshifts of the variable sources (average 
redshift $\sim$0.95) and the rest-frame monochromatic 24 $\mu$m luminosity. The  mean value of 
rest-frame $\log (\nu L_{24\mu{m}} / L_{\odot})$ is 10.5 for both the long-term
  and the short-term variable sources. For those
  candidates satisfying the \cite{La} AGN selection criteria the mean
  values are $\log (\nu L_{24\mu{m}} / L_{\odot}) =10.7$ for both, the
  long-term and the short-term 
  variable candidates. Conversely, the candidates not
  satisfying the \cite{La} criteria have mean values of $\log (\nu L_{24\mu{m}} / L_{sun})= 10.3$ and 
  10.4 for the long-term and the short-term variable candidates, respectively.

We studied the radio properties of the variable sources. Only 9 sources have radio data. Of them, 44\% present 
radio excess, defined as $q=log(f_{24~\mu{\rm m}}/f_{\rm 1.4~GHz})$ which might be an indication of AGN activity (see \cite{Appleton04}).

We also studied the X-ray properties and obtained that the 30\% of the 24 $\mu$m variable sourced are 
detected in X-rays in the central part of the E-CDFS (covered by \cite{Xue}). The X-ray $0.5-8\,$keV luminosities 
are ranging from $ \sim  1\times 10^{40}\,{\rm erg \,s}^{-1}$ to $\sim 1\times 10^{44}\,{\rm erg \,s}^{-1}$. 4\% of 
the X-ray sources satisfying the properties of our parent MIPS 24 $\mu$m catalogs are variable at 24 $\mu$m on the
timescales probed by our study. Assuming that deep X-ray exposures provide the majority of the AGN, the 24 $\mu$m variable 
sources not detected in X-ray would only account $\sim$13\% of the total AGN population.

Finally, the compared our variable sources with sources selected as AGN by other criteria. We 
found $\sim$56\% of the variable 24 $\mu$ sources in the E-CDFS would be identified as AGN by other methods. Table \ref{fractions-candidates} 
summarizes this comparison.

\begin{table}
  \caption{Summary of fractions of  MIPS $24\,\mu$m variable sources 
    selected as AGN by other criteria.}
   \begin{minipage}{150mm}

    \begin{tabular}{@{}ccccccccc}
      
      \hline\hline
      &\tiny No. &\tiny X-ray$^1$&\tiny radio$^2$ &\tiny  other AGN$^3$  &\tiny IR$^4$ &\tiny Combined$^5$ &\tiny \cite{La}$^6$  &\tiny  Combined$^7$ \\
      &\tiny variable& &\tiny excess&\tiny catalogs&\tiny  power law& &\tiny criteria&\tiny criteria\\
      &\tiny sources &\tiny No. (\%)&\tiny No. (\%)&\tiny No. (\%)&\tiny No. (\%)&\tiny No. (\%) &\tiny No. (\%)&\tiny No.(\%)\\
      \hline
      \multicolumn{9}{c}{\tiny Long-term variable sources}\\
      \hline \hline
     \tiny In  the E-CDFS &\tiny 28 &\tiny 7 (25) &\tiny 2 (7) &\tiny 4 (14) &\tiny 1 (4) &\tiny 8 (29) &\tiny 12 (43) &\tiny 17 (61)\\
      \tiny Outside the E-CDFS &\tiny 11 &\tiny 0 (0) &\tiny 0 (0) &\tiny 0 (0) &\tiny 2 (18) &\tiny 2 (18) &\tiny 5 (45) &\tiny 5 (45)\\
     \tiny All &\tiny 39 &\tiny 7 (18) &\tiny 2 (5) &\tiny 4 (10) &\tiny 3 (8) &\tiny 10 (26) &\tiny 17 (44) &\tiny 22 (56) \\
     \hline
     \multicolumn{9}{c}{\tiny Short-term variable sources}\\
      \hline \hline
      \tiny In the E-CDFS &\tiny 33 &\tiny 4 (12) &\tiny 1 (3) &\tiny 3 (9) &\tiny 0 (0) &\tiny 5 (15) &\tiny 14 (42) &\tiny 17 (52)\\
      \tiny Outside the E-CDFS&\tiny 22 &\tiny 1 (5) &\tiny 1 (5) &\tiny 1 (5) &\tiny 1 (5) &\tiny 2 (9) &\tiny 5 (23) &\tiny 5 (23)\\
     \tiny All&\tiny 55 &\tiny 5 (9) &\tiny 2 (4) &\tiny 4 (7) &\tiny 1 (2) &\tiny 7 (13) &\tiny 19 (35) &\tiny 22 (40)\\
      \hline

    \end{tabular}
    
\tiny $^1$ Variable MIPS 24 $\mu$m sources detected in X-rays. \tiny $^2$ Variable MIPS 24 $\mu$m sources with radio excess. \tiny $^3$ Variable 
MIPS 24 $\mu$m sources in other AGN catalogs. \tiny $^4$ Variable MIPS 24 $\mu$m sources detected as
IR power-law AGN. \tiny $^5$ Combined 1$^{st}$, 2$^{nd}$, 3$^{rd}$, and 4$^{th}$ criteria. \tiny $^6$ Variable MIPS 24 $\mu$m sources 
satisfying the \cite{La} criteria. \tiny $^7$ All the criteria combined.

  \end{minipage} 
 \label{fractions-candidates}    
\end{table}

\section{Summary and conclusions}

We used a $\chi^2$ method to select long-term (years) and short-term (days) variable sources at 24 $\mu$m using 
deep {\em Spitzer}/MIPS imaging data from 7 epochs (2004-2007) in the GOODS-South field. We found 39 long-term and 
55 short-term variable sources. Of them, 28 long-term and 33 short-term variable sources are located in the E-CDFS. The
average redshift is $\sim$0.95. We are therefore probing typically variable emission at 12 $\mu$m rest-frame. The contribution
of the AGN to the 24 $\mu$ emission is low, which probably implies that they are low-luminosity AGN. 30\% of the variable
sources are detected in X-ray in the central part of the E-CDFS. Sources without X-ray detection are $\leq$13\% of the total 
AGN population in the central part of the E-CDFS. Approximately 56\% of the variable sources in the E-CDFS would be identified
as AGN by other methods. Therefore, MIPS 24 $\mu$m variability provides a new method to identify AGN in cosmological fields. See \cite{Judit} for more details.

%
%
\small  
%
\section*{Acknowledgments}   
%
J.G.-G., A.A.-H., and A.H.-C. acknowledge support from the Augusto
G. Linares research program of the Universidad de Cantabria and from
the Spanish Plan Nacional through grant AYA2012-31447. P.G.P.-G. acknowledges support from MINECO grant AYA2012-31277.

%

%

\begin{thebibliography}{}
\small
%

\bibitem{Almu}{Alonso-Herrero A. et al., 2006, ApJ, 640, 167}
\bibitem{Antonucci1993}{Antonucci R., 1993, ARA\&A, 31, 473}
\bibitem{Appleton04}{Appleton P. N. et al., 2004, ApJS, 157, 147}
\bibitem{BA}{Bertin E., \& Arnouts S., 1996, A\&AS, 117, 393}
\bibitem{Do}{Donley J. L. et al., 2012, ApJ, 748, 142}
\bibitem{Judit}{Garc\'ia-Gonz\'alez J., Alonso-Herrero A., Per\'ez-Gonz\'alez P. G., Hern\'an-Caballero A., Sarajedini V. L., Villar V., 2014, MNRAS, in press (arXiv:1410.6011)}
\bibitem{Hickox2014}{Hickox R. C., Mullaney J. R., Alexander D. M., Chen C. J., Civano F. M., Goulding A. D., Hainline K. N, 2014, ApJ, 782, 9}
\bibitem{HK}{H\"onig S. F., \& Kishimoto M., 2011, A\&A, 534, 121}
\bibitem{HB}{Hopkins A. M., \& Beacom J. F., 2006, ApJ, 651, 142}
\bibitem{Ko}{Kozlowski S. et al., 2010, ApJ, 716, 530}
\bibitem{La}{Lacy M. et al., 2004, ApJS, 154, L166}
\bibitem{La13}{Lacy M. et al., 2013, ApJS, 208, 24}
\bibitem{Pe}{Pereyra N. A., Vanden Berk D. E., Turnshek D. A., Hillier D. J., Wilhite B. C., Kron R. G., Schneider D. P., Brinkmann J., 2006, ApJ, 642, 87}
\bibitem{Perez-Gonzalez2005}{P\'erez -Gonz\'alez P. G. et al., 2005, ApJ, 630, 82}
\bibitem{Perez-Gonzalez2008}{P\'erez -Gonz\'alez P. G. et al., 2008, ApJ, 675, 234}
\bibitem{Rie}{Rieke G. H. et al., 2004, ApJS, 154, 25}
\bibitem{Ste}{Stern D. et al., 2005, ApJ, 631, 163}
\bibitem{Tre}{Trevese D., Kron R. G., Majewski S. R., Bershady M. A., Koo D. C., 1994, ApJ, 433, 494}
\bibitem{UMU}{Ulrich M., Maraschi L., Urry C. M., 1997, ARA\&A, 35, 445}
\bibitem{We}{Werner M. W. et al., 2004, ApJS, 154, 1}
\bibitem{Xue}{Xue Y. Q. et al., 2011, ApJS, 195, 10}










%
%
\end{thebibliography}
\end{document}